\documentclass{ws-procs9x6}

\def\be{\begin{equation}}
\def\ee{\end{equation}}
\def\bea{\begin{eqnarray}}
\def\eea{\end{eqnarray}}
\def\ba{\begin{array}}
\def\ea{\end{array}}

\begin{document}

\title{NONPERTURBATIVE QUARK SEA ASYMMETRIES}

\author{H. DAHIYA$^*$ and N. SHARMA}

\address{Department of Physics, Dr. B.R. Ambedkar National
Institute of Technology,\\ Jalandhar, 144011, India\\ $^*$E-mail:
harleen\_dahiya@yahoo.com}

\begin{abstract}
The effects of  nonperturbatively  generated ``quark sea'' have
been investigated to determine the flavor structure of the octet
baryons. The chiral constituent quark model ($\chi$CQM), which is
known to provide a satisfactory explanation of the proton spin and
related issues in the nonperturbative regime, is able to explain
the qualitative generation of the requisite amount of quark sea.
The importance of quark sea has been studied at different values
of the Bjorken scaling variable $x$ by including it
phenomenologically in the sea quark distribution functions. The
results for the quark sea asymmetries like $\bar d(x)-\bar u(x)$,
$\bar d(x)/\bar u(x)$ and Gottfried integral for the octet baryons
strengthen the significance of quark sea at lower values of $x$.

\end{abstract}

\keywords{Chiral symmetry breaking; Nonperturbative regime of QCD;
Octet baryons.}

\bodymatter

\section{Introduction}

After the first direct evidence for the point-like constituents in
the nucleon\cite{point-like}, identified as the valence quarks
with spin-1/2 in the naive constituent quark model
(NQM)\cite{dgg,Isgur,yaouanc}, a lot of experiments have been
conducted to probe the structure of the proton in the deep
inelastic scattering (DIS) experiments. Surprisingly, the DIS
results in the early 80's\cite{emc} indicated that the valence
quarks of the proton carry only about 30\% of its spin and is
referred to as the ``proton spin crisis'' in the NQM. These
results provided the first evidence for the proton being composed
of three valence quarks surrounded by an indistinct sea of
quark-antiquark pairs (henceforth referred to as the ``quark
sea'').  In the present day, the study of the composition of
hadrons can be said to be primarily the study of the quark sea and
gluons and is considered as one of the active areas in hadronic
physics.

The conventional expectation that the quark sea perhaps can be
obtained through the perturbative production of the
quark-antiquark pairs by gluons produces nearly equal numbers of
$\bar u$ and $\bar d$. Until early 90's a symmetric sea w.r.t.
$\bar u$ and $\bar d$ was assumed, however, the famous New Muon
Collaboration in 1991\cite{nmc} established the quark sea
asymmetry of the unpolarized quarks in the case of nucleon by
measuring $\bar d-\bar u$ giving first clear evidence for the
nonperturbative origin of the quark sea. This was later confirmed
by the Drell-Yan experiments\cite{baldit} which measured  a large
quark sea asymmetry ratio $\bar d/\bar u$ reminding us that the
study of the quark sea is intrinsically a nonperturbative
phenomena and it is still a big challenge to perform these
calculations from the first principles of QCD.

One approach to account for the observed quark sea asymmetry is
the pion cloud mechanism\cite{pioncloud} where the quark sea is
believed to originate from process such as virtual pion
production. It is suggested that in the deep inelastic
lepton-nucleon scattering, the lepton probe also scatters off the
pion cloud surrounding the target proton. The $\pi^+ (\bar d u)$
cloud, dominant in the process $p \rightarrow \pi^+ n$, leads to
an excess of $\bar d$ sea. However, this effect should be
significantly reduced by the emissions such as $p \rightarrow
{\Delta}^{++} + \pi^-$ with  $\pi^- (\bar u d)$ cloud. Therefore,
the pion cloud idea is not able to explain the significant $\bar d
> \bar u$ asymmetry. This approach can be improved upon by
adopting a mechanism which operates in the {\it interior} of the
hadron.

The chiral constituent quark model ($\chi$CQM)\cite{manohar} can
yield an adequate description of the quark sea generation through
the chiral fluctuations. The basic idea is based on the
possibility that chiral symmetry breaking takes place at a
distance scale much smaller than the confinement scale. In this
region, the effective degrees of freedom are the valence quarks
and the internal Goldstone bosons (GBs) which are coupled to the
valence quarks\cite{eichten,cheng,song} allowing a simple and
intuitive method to investigate the principle features of the
hadron structure. In the case of spin dependent quantities, the
$\chi$CQM is not only successful in giving a satisfactory
explanation of ``proton spin crisis''\cite{eichten,hd} but is also
able to account for the baryon magnetic moments\cite{hdorbit} and
hyperon $\beta-$decay parameters\cite{cheng,johan,ns}. However, in
the case of quark distribution functions, the latest developments
by the NuSea (E866)\cite{e866} and HERMES\cite{hermes} to
determine the variation of the sea-antiquark ratio $\bar d(x)/\bar
u(x)$ and the difference $\bar d(x)-\bar u(x)$ with Bjorken
scaling variable $x$ have renewed considerable interest in the
quark sea asymmetries.

Recently, there has been substantial theoretical  progress to take
into account the effects of quark sea in determining the flavor
structure of the baryons and the question of sea asymmetry has
been investigated  by several authors using various
phenomenological models. Calculations have been carried out in the
meson cloud models\cite{mesoncloud}, chiral quark-soliton
model\cite{cqsm}, effective chiral quark model\cite{eccm},
statistical models\cite{stat}, bag model\cite{bag}, model for
parton densities\cite{alwall}, radiative parton model\cite{reya}
etc.. However, the inclusion of $x-$dependence has not yet been
successfully included in the quark distribution functions.
Therefore, pending further experiments, it would be interesting to
examine the flavor structure of the octet baryons at low energy,
thereby giving vital clues to the nonperturbative effects of QCD.
The study of $x-$dependence in the quark distribution functions
becomes particularly interesting for the $\chi$CQM where the
effects of quark sea and valence quarks can separately be
calculated.

The purpose of the present communication is to determine the sea
quark distribution functions and their asymmetries in the octet
baryons by phenomenologically incorporating $x-$dependence in the
$\chi$CQM. The extent of contributions coming from the different
sea quarks for the octet baryons can also be compared. To
understand the relation of the Bjorken scaling variable and quark
sea, it would be significant to study its implications in the
region $x<0.3$ which is a relatively clean region to test the
quark sea structure as well as to estimate their structure
functions and related quantities\cite{cdhs}.

\section{Chiral Constituent Quark Model}

The key to understand the ``proton spin crisis'', in the $\chi$CQM
formalism, is the fluctuation process\cite{cheng} $q^{\pm}
\rightarrow {\rm GB} + q^{' \mp} \rightarrow  (q \bar q^{'})
+q^{'\mp},$ where 
$q \bar q^{'} +q^{'}$ constitute the ``quark
sea''\cite{cheng,song,hd,johan}. The effective Lagrangian
describing interaction between quarks and a nonet of GBs,  can be
expressed as ${\cal L}= g_8 {\bf \bar
q}\left(\Phi+\zeta\frac{\eta'}{\sqrt 3}I \right) {\bf q}=g_8 {\bf
\bar q}\left(\Phi' \right) {\bf q},$ where $\zeta=g_1/g_8$, $g_1$
and $g_8$ are the coupling constants for the singlet and octet
GBs, respectively, $I$ is the $3\times 3$ identity matrix. In
terms of the SU(3) and axial U(1) symmetry breaking parameters,
introduced by considering $M_s > M_{u,d}$, $M_{K,\eta}> M_{\pi}$
and $M_{\eta^{'}} > M_{K,\eta}$\cite{cheng,song,johan}, the GB
field can be expressed as \bea \Phi' = \left( \ba{ccc}
\frac{\pi^0}{\sqrt 2} + \beta\frac{\eta}{\sqrt 6} +
\zeta\frac{\eta^{'}}{\sqrt 3} & \pi^+& \alpha K^+ \\ \pi^- &
-\frac{\pi^0}{\sqrt 2} +\beta \frac{\eta}{\sqrt 6}
+\zeta\frac{\eta^{'}}{\sqrt 3}  &  \alpha K^0
\\\alpha K^-  & \alpha \bar{K}^0  &  -\beta \frac{2\eta}{\sqrt 6}
+\zeta\frac{\eta^{'}}{\sqrt 3} \ea \right)\,. \eea The parameter
$a(=|g_8|^2$) denotes the probability of chiral fluctuation  $u(d)
\rightarrow d(u) + \pi^{+(-)}$, whereas $\alpha^2 a$, $\beta^2 a$
and $\zeta^2 a$ respectively denote the probabilities of
fluctuations $u(d) \rightarrow s + K^{-(0)}$, $u(d,s) \rightarrow
u(d,s) + \eta$, and $u(d,s) \rightarrow u(d,s) + \eta^{'}$.

For the sake of simplification, the GB field can also be expressed
in terms of the quark contents of the GBs and is expressed as \be
\left( \ba{ccc} \phi_{uu} u \bar u+ \phi_{ud} d \bar d +\phi_{us}
s \bar s& \varphi_{ud} u \bar d & \varphi_{us} u \bar s
\\ \varphi_{du} d \bar u & \phi_{du}u \bar u+ \phi_{dd} d \bar d
+\phi_{ds} s \bar s & \varphi_{ds} d \bar s
\\ \varphi_{su} s \bar u & \phi_{sd} s \bar d & \phi_{su} u \bar u
+ \phi_{sd} d \bar d +\phi_{ss} s \bar s \\ \ea \right), \ee where
\bea \phi_{uu} &=& \phi_{dd}= \frac{1}{2} +\frac{\beta}{6} +
\frac{\zeta}{3},\phi_{ss} =\frac{2\beta}{3} + \frac{\zeta}{3},
\phi_{us} = \phi_{ds}= \phi_{su}=\phi_{sd}= -\frac{\beta}{3} +
\frac{\zeta}{3}, \nonumber \\ \phi_{du} &=& \phi_{ud}=
-\frac{1}{2} +\frac{\beta}{6} + \frac{\zeta}{3}, \varphi_{ud} =
\varphi_{du} = 1,\varphi_{us} = \varphi_{ds}=\varphi_{su}=
\varphi_{sd}= \alpha.  \eea The quark sea content of the baryon
can be calculated in $\chi$CQM by substituting for every
constituent quark $q \to \sum P_q q + |\psi(q)|^2$, where $\sum
P_q$ is the transition probability of the emission of a GB from
any of the $q$ quark and $|\psi(q)|^2$ is the transition
probability of the $q$  quark. The flavor structure for the baryon
of the type $B(xxy)$ is expressed as $2 P_x x + P_y y + 2
|\psi(x)|^2 + |\psi(y)|^2$ and for the type $B(xyz)$ it is
expressed as $ P_x x + P_y y + P_z z+ |\psi(x)|^2 + |\psi(y)|^2+
|\psi(z)|^2$, where $x,y,z=u,d,s$.

\section{Quark Distribution Functions}

There are no simple or straightforward rules which could allow
incorporation of $x-$dependence in $\chi$CQM. To this end, instead
of using an {\it ab initio} approach, we have phenomenologically
incorporated the $x-$dependence getting clues from Eichten {\it et
al.}\cite{eichten}, Isgur  {\it et al.}\cite{Isgur} and Le Yaouanc
{\it et al.}\cite{yaouanc}. The $x-$dependent sea quark
distribution functions can be now expressed as $\bar u^B(x)=\bar
u^B{(1-x)}^{10}$, $\bar d^B(x)=d^B{(1-x)}^{7}$, $\bar
s^B(x)=s^B{(1-x)}^{8}$ which together with the valence quark
distribution functions give the flavor structure of the baryon as
\be q^B(x)=q^B_{{\rm val}}(x)+\bar q^B(x)\,, \label{totalquark}\ee
where $q=u,d,s$. Using the sea quark distribution functions as
explained above, the quark sea asymmetries $\bar u(x) - \bar d(x)$
and $\bar d(x)/\bar u(x)$ can also be calculated at different $x$
values. We have already discussed the inclusion of $x$-dependence
in detail and compared our results with the experimental data for
the case of nucleon in Ref.\cite{xdep}. In the present
communication however, we have extended our calculations to the
case of other octet baryons for which experimental data is not yet
available.

The $x-$dependence of the structure functions $F_1$ and $F_2$ can
be calculated from \be F_2^B(x) = x \sum_{u,d,s} e^2_{q}[q^B(x) +
\bar q^B(x)] \,, ~~~ F_1^B(x) = \frac{1}{2x}F_2^B(x) \,, \ee where
$e_q$ is the charge of the quark $q$ ($e_u=\frac{2}{3}$ and $e_d=
e_s=-\frac{1}{3}$). In terms of the quark distribution functions,
the structure function $F_2$ for any baryon can be expressed as
\be F^B_2(x)=\frac{4}{9} x(u^B(x)+ \bar u^B(x)) +\frac{1}{9}
x(d^B(x)+ \bar d^B(x)+ s^B(x)+\bar s^B(x))\,. \ee Several
important quantities can be obtained from the structure functions
of different isospin multiplets. For example, for the case of
proton and neutron we have \be \frac{F^p_2(x) -F^n_2(x)}{x} =
\frac{4}{9}\left(u_{\rm val}^p(x) - u_{\rm val}^n(x) + 2\bar
u^p(x) - 2\bar u^n(x) \right) +\frac{1}{9} \left( d_{\rm
val}^p(x)\right. \nonumber \ee \be \left. + s_{\rm val}^p(x) -
d_{\rm val}^n(x)- s_{\rm val}^n(x) + 2\bar d^p(x)+ 2 \bar s^p(x) -
2 \bar d^n(x)- 2 \bar s^n(x) \right). \label{ig} \ee

The Gottfried integral $I^{p n}_G$\cite{gsr} can be expressed in
terms of the sea quarks as follows \be I^{p n}_G = \int_0^1
{\frac{F^p_2(x) -F^n_2(x)}{x}} dx = \frac{1}{3} + \frac{2}{3}
\int_0^1 \left[ \bar u^p(x)- \bar d^p(x) \right] dx\,,
\label{ipn}\ee where we have used the following normalization
conditions \[ \int_0^1 u_{\rm val}^p(x)dx = 2 \,,~~ \int_0^1
d_{\rm val}^p(x)dx=1 \,, \int_0^1 s_{\rm val}^p(x)dx=0 \,, \]
\[\int_0^1 u_{\rm val}^n(x)dx=1 \,, \int_0^1 d_{\rm val}^n(x)dx=2
\,, \int_0^1 s_{\rm val}^n(x)dx=0 \,, \]

\be \int_0^1 \bar d^n(\bar u^n)(x)dx= \int_0^1 \bar u^p (\bar
d^p)(x) dx \,,~~ \int_0^1 \bar s^n(x)dx= \int_0^1 \bar s^p(x) dx
\,. \ee
Similarly, for the case of other octet baryons 
the following normalization conditions  \bea \int_0^1 u_{\rm
val}^{\Sigma^+}(x)dx = 2 \,,~~~ \int_0^1 d_{\rm
val}^{\Sigma^+}(x)dx = 0 \,,~~~ \int_0^1 s_{\rm val}^{\Sigma^+} dx
= 1 \,, \nonumber \\ \int_0^1 u_{\rm val}^{\Sigma^0}(x)dx = 1
\,,~~~ \int_0^1 d_{\rm val}^{\Sigma^0}(x) dx=1 \,,~~~ \int_0^1
s_{\rm val}^{\Sigma^0}(x) dx = 1 \,, \nonumber\\ \int_0^1 u_{\rm
val}^{\Xi^{0}}(x) dx = 1 \,,~~~ \int_0^1 d_{\rm val}^{\Xi^{0}}(x)
dx = 0 \,,~~~ \int_0^1 s_{\rm val}^{\Xi^{0}}(x) dx = 2 \,,
\label{norm-s} \eea lead to the other Gottfried integrals in terms
of the sea quarks \bea I^{\Sigma^+ \Sigma^0}_G &=& \frac{1}{3} +
\frac{2}{9} \int_0^1 \left[ 4 \bar u^{\Sigma^+}(x)+ \bar
d^{\Sigma^+}(x) - 4 \bar u^{\Sigma^0}(x) -\bar d^{\Sigma^0}(x)
\right] dx \,,\nonumber\\ I^{\Sigma^0 \Sigma^-}_G  &=& \frac{1}{3}
+ \frac{2}{9} \int_0^1 \left[ 4 \bar u^{\Sigma^0}(x)+ \bar
d^{\Sigma^0}(x) - 4 \bar d^{\Sigma^+}(x) - \bar u^{\Sigma^+}(x)
\right] dx\,,\nonumber\\  I^{\Xi^{0} \Xi^{-}}_G &=& \frac{1}{3} +
\frac{2}{3} \int_0^1 \left[\bar u^{\Xi^0}(x) - \bar d^{\Xi^0}(x)
\right] dx\,. \label{ioctet}\eea It is clear from Eqs. (\ref{ipn})
and (\ref{ioctet}), the flavor symmetric sea leads to the
Gottfried sum rule $I_G=\frac{1}{3}$ with $\bar u^B$=$\bar d^B$.

\section{Results and Discussion}
After having detailed the contribution of the quark sea and the
various asymmetries in the octet baryons of different quark
structure, we now discuss the variation of these quantities with
the Bjorken  variable $x$. For the numerical calculation of the
sea quark distribution functions of the octet baryons, we have
used the same set of input parameters as detailed in our earlier
calculations\cite{hd,ns,xdep,pdg}. One can easily find out that

\[\bar d^{p}(x)>\bar u^{p}(x)>\bar s^{p}(x),\]
\[\bar d^{\Sigma^+}(x)>\bar u^{\Sigma^+}(x)\approx\bar s^{\Sigma^+}(x),\]
\[\bar d^{\Sigma^0}(x)>\bar u^{\Sigma^0}(x)>\bar s^{\Sigma^0}(x),\]
\[\bar d^{\Xi^0}(x)>\bar u^{\Xi^0}(x)>\bar s^{\Xi^0}(x),\]
showing a clear quark sea asymmetry as observed in the DIS
experiments\cite{nmc,e866,hermes}. These distributions clearly
indicate that our results pertaining to the quark sea asymmetry
seem to be well in line with the expected results. As already
mentioned in the introduction, the sea quarks do not contribute at
higher values of $x$, therefore in this region, the contributions
should be completely dominated by the valence quarks. The
difference between the various sea distributions is observed to be
maximum at $x\approx0.1$. As the value of $x$ increases, the
difference between the sea contributions decreases in all the
cases which is in line with the observations of other
models\cite{cqsm,stat,bag,alwall}.

The general aspects of the variation of the magnitudes of the sea
quark distribution functions $\bar u(x)$, $\bar d(x)$ and $\bar
s(x)$ for the octet baryons are able to explain some of the well
known experimentally measurable quantities, for example, $\bar
d^{B}(x)-\bar u^{B}(x)$, $\bar d^{B}(x)/\bar u^{B}(x)$ and the
Gottfried integral. These quantities not only provide important
constraint on a model that attempts to describe the origin of the
quark sea but also provide a direct determination of the presence
of significant amount of quark sea in the low $x$ region. When $x$
is small, $\bar d^{B}(x)-\bar u^{B}(x)$ asymmetries are large
implying the dominance of sea quarks in the low $x$ region. In
fact, the sea quarks dominate only in the region where $x$ is
smaller than 0.3. At the values $x>0.3$, $\bar d-\bar u$ tends to
0 implying that there are no sea quarks in this region. The
contribution of the quark sea in the case of $\Sigma^0$ is
particularly interesting because of its flavor structure which has
equal numbers of $u$, $d$ and $s$ quarks in its valence structure.
Unlike the other octet baryons, where the $\bar d(x)-\bar u(x)$
asymmetry decreases continuously with the $x$ values, the
asymmetry in this case first increases and then for values of
$x>0.1$ it decreases. However, it is interesting to observe the
the asymmetry peak in this case which matches with our other
predictions where the contribution of the quark sea is maximum at
$x\approx0.1$

A measurement of the Gottfried integral\cite{nmc,e866} for the
case of nucleon has shown a clear violation of Gottfried sum rule
from $\frac{1}{3}$  which can find its explanation in a global
quark sea asymmetry $ \int_0^1 (\bar d(x) -\bar u(x))dx$.
Similarly, for the case of $\Sigma^{+}$, $\Sigma^{0}$, and
$\Xi^0$, the Gottfried sum rules should read $I^{\Sigma^+
\Sigma^0}_G=\frac{1}{3}$, $I^{\Sigma^0 \Sigma^-}_G=\frac{1}{3}$
and $I^{\Xi^0 \Xi^-}_G=\frac{1}{3}$ if the quark sea was
symmetric. However, due to the $\bar d(x)-\bar u(x)$ asymmetry in
the case of octet baryons, a lower value of the Gottfried
integrals is obtained.  In the case of nucleon the results are in
good agreement with the experimental data\cite{e866} as already
presented in\cite{xdep}. The quality of numerical agreement in the
other cases can be assessed only after the data gets refined.
Further, this phenomenological analysis strongly suggests an
important role for the quark sea at low value of $x$. New
experiments aimed at measuring the flavor content of the other
octet baryons are needed for profound understanding of the
nonperturbative properties of QCD.

\section{Summary and Conclusions}
To summarize, in order to investigate the effects of ``quark
sea'', we have calculated the sea quark distribution functions for
the octet baryons  in the chiral constituent quark model
($\chi$CQM). The Bjorken scaling variable $x$ has been
incorporated phenomenologically to enlarge the scope of model and
to understand the range of $x$ where quark sea effects are
important. Implications of the quark sea have also been studied to
estimate the quark sea asymmetries like $\bar d(x)-\bar u(x)$,
$\bar d(x)/\bar u(x)$ and Gottfried integral. The results justify
our conclusion regarding  the importance of quark sea  at small
values of $x$.

In conclusion, the results obtained for the quark distribution
functions reinforce our conclusion that $\chi$CQM is able to
generate qualitatively as well as quantitatively the requisite
amount of quark sea. This can perhaps be substantiated by a
measurement of the quark distribution functions of the other octet
baryons.

\section*{Acknowledgments} H.D. would like to thank the
organizers, Gribov80 Workshop held at ICTP and Department of
Science and Technology, Government of India, for financial
support.

\end{document}